\documentstyle[12pt,aasms4]{article}

\def\kms{km~s$^{-1}$}

\received{}
\accepted{}
\journalid{}{}
\articleid{}{}
\lefthead{Hatano et al.}
\righthead{Type Ia SN 1994D}

\begin{document}
\title {On the High--Velocity Ejecta of the Type Ia Supernova 1994D}

\author {Kazuhito Hatano{\altaffilmark{1}}, David
Branch{\altaffilmark{1}}, {Adam Fisher\altaffilmark{1}},
{E.~Baron\altaffilmark{1}}, and {A.~V.~Filippenko\altaffilmark{2}}}

\altaffiltext{1}{Department of Physics and Astronomy, University of
Oklahoma, Norman, Oklahoma 73019, USA}

\altaffiltext{2}{Department of Astronomy, University of California,
Berkeley, CA 94720--3411}

\begin{abstract}

Synthetic spectra generated with the parameterized supernova
synthetic--spectrum code SYNOW are compared to spectra of the Type~Ia
SN~1994D that were obtained before the time of maximum brightness.
Evidence is found for the presence of two--component Fe~II and Ca~II
features, forming in high velocity ($\ge 20,000$ \kms) and lower
velocity ($\le 16,000$ \kms) matter.  Possible interpretations of
these spectral splits, and implications for using early--time spectra
of SNe~Ia to probe the metallicity of the progenitor white dwarf and
the nature of the nuclear burning front in the outer layers of the
explosion, are discussed.
      
\end{abstract}


\section{Introduction}

SN~1994D, in the Virgo cluster galaxy NGC~4526, was discovered two
weeks before the time of maximum brightness (Treffers et~al. 1994) and
became the best observed Type~Ia supernova (Richmond et~al. 1995;
Patat et~al. 1996; Meikle et~al. 1996; Vacca \& Leibundgut 1996;
Filippenko 1997a,b). The high--quality early--time spectra of
SN~1994D, beginning 12 days before maximum brightness, present us with
an opportunity to look for spectral lines produced by ``primordial''
matter, i.e., heavy elements that were already present in the
progenitor white dwarf before it exploded, as well as to probe the
nature of the nuclear burning front in the outer layers of the ejected
matter.

In this paper we report some results of a ``direct analysis'' of
photospheric--phase spectra of SN~1994D using the parameterized
supernova spectrum--synthesis code SYNOW (Fisher et~al. 1997, 1999;
Millard et~al. 1999; Fisher 1999).  Here we concentrate mainly on
spectra obtained before maximum light. An analysis of the
post--maximum photospheric--phase spectra will be presented in another
paper.

\section{Previous Studies of SN~1994D Spectra}

Numerous optical spectra of SN~1994D have been published by Patat
et~al. (1996), Meikle et~al. (1996), and Filippenko (1997a,b).  Patat
et~al. emphasized that in spite of some photometric peculiarities such
as being unusually blue, SN~1994D had the spectral evolution of a
normal SN~Ia, especially like that of SN~1992A (Kirshner
et~al. 1993). Meikle et~al. also emphasized the spectral resemblance
to SN~1992A, and presented some near--infrared spectra.

H\"oflich (1995) calculated light curves and detailed
non--local--thermodynamic--equilibrium (NLTE) spectra for a series of
delayed--detonation hydrodynamical models and found that one
particular model --- M36, which contained 0.60 $M_\odot$ of freshly
synthesized $^{56}$Ni --- gave a satisfactory representation of the
light curves and spectra of SN~1994D.  Synthetic spectra for model M36
were compared to observed SN~1994D spectra at epochs of $-10, -5, 0,
11$, and 14 days.  (We cite spectrum epochs in days with respect to
the date of maximum brightness in the $B$ band, 1994 March~21 UT
[Richmond et~al. 1995].)  The good agreement between the calculated
and observed spectra and light curves indicated that the composition
structure of SN~1994D resembled that of model M36 at least in a
general way.

Patat et~al. (1996) compared their observed $-4$ day spectrum to a
Monte Carlo synthetic spectrum calculated for the carbon--deflagration
model W7 (Nomoto, Thielemann, \& Yokoi 1994), which has a composition
in its outer layers that differs significantly from model that of M36.
This calculated spectrum also agreed reasonably well with the observed
spectrum.

Meikle et~al. (1996) concentrated on identifying an infrared P~Cygni
feature that appeared in their pre--maximum--light spectra.  They
considered He~I $\lambda10830$ and Mg~II $\lambda10926$, and found
that in parameterized synthetic spectra either transition could be made
to give a reasonable fit, but then other transitions of these ions
produced discrepancies elsewhere in the spectrum.  Mazzali \& Lucy
(1998) also focused on identifying the infrared feature.  They found
that either He~I or Mg~II could be made to give a reasonable fit, but
that neither could explain the rapid disappearance of the observed
feature after the time of maximum brightness.  Wheeler et~al. (1998)
favored the Mg~II identification and concluded that within the context
of delayed--detonation models the feature blueshift provides a
sensitive diagnostic of the density at which the deflagration switches
to a detonation.

\section{Spectrum Synthesis Procedure}

We have been using the fast, parameterized, supernova
spectrum--synthesis code SYNOW to make a direct analysis of
photospheric--phase spectra of SN~1994D.  The goal has been to
establish line identifications and intervals of ejection velocity
within which the presence of lines of various ions are detected,
without adopting any particular hydrodynamical model.  In our work on
SN~1994D we have made use of the results of Hatano et~al. (1999), who
presented plots of LTE Sobolev line optical depths versus temperature
for six different compositions that might be expected to be
encountered in supernovae, and also presented SYNOW optical spectra
for 45 individual ions that can be regarded as candidates for
producing identifiable spectral features in supernovae.  (Electronic
data for the Hatano et~al. [1999] paper, now extended to include the
near infrared, can be obtained at
www.nhn.ou.edu/$\sim$baron/papers.html).

For comparison with each observed spectrum, we have calculated many
synthetic spectra with various values of the fitting parameters.
These include $T_{bb}$, the temperature of the underlying blackbody
continuum; $T_{exc}$, the excitation temperature; and $v_{phot}$, the
velocity of matter at the photosphere. For each ion that is
introduced, the optical depth of a reference line also is a fitting
parameter, with the optical depths of the other lines of the ion being
calculated for LTE excitation at $T_{exc}$.  We also can introduce
restrictions on the velocity interval within which each ion is
present; when the minimum velocity assigned to an ion is greater than
the velocity at the photosphere, the line is said to be detached from
the photosphere.  The radial dependence of all of the line optical
depths is taken to be exponential with {\sl e}--folding velocity $v_e
= 3000$ km s$^{-1}$, and the line source function is taken to be that
of resonance scattering.  All of the adopted fitting parameters are
given in Tables 1-3.  The most interesting parameters are $v_{phot}$,
which as expected is found to decrease with time, and the individual
ion velocity restrictions, which constrain the composition structure.

\section{Results}

\subsection{Twelve Days Before Maximum}

The $-12$ day observed spectrum appears in both the upper and lower
panels of Figure~1.  The upper panel also contains a SYNOW synthetic
spectrum based on what could be called the conventional interpretation
of early--time SN~Ia spectra (Filippenko 1997 and references therein).
The adopted value of $v_{phot}$ is 15,000 \kms. The ions that are
certainly required to account for certain spectral features are Ca~II,
Si~II, S~II, and Fe~II.  In an attempt to improve the fit we also have
introduced weaker contributions from C~II, Na~I, Mg~II, Si~III,
Fe~III, Co~II, and Ni~II, some with minimum and maximum velocities as
listed in Table~1.  (Ions for which no minimum velocity is listed are
undetached, i.e., their listed optical depths refer to the velocity at
the photosphere, 15,000 \kms.)  In spite of having a considerable
number of free parameters at our disposal, we are left with two
serious discrepancies: the observed absorption minima near 4300 and
4700~\AA\ are not accounted for.  (The discrepancy from 3900~\AA\ to
4200~\AA\ is not very troubling because SYNOW spectra often are
underblanketed in the blue due to the lack of weak lines of unused
ions.)  We have been unable to remove the 4300 and 4700~\AA\
discrepancies by introducing any additional ions that would be
plausible under these circumstances, according to the LTE calculations
of Hatano et~al. (1999).

The lower panel of Figure~1 shows what we consider to be the most
plausible way to improve the fit.  The only difference in the
synthetic spectrum is that we have introduced a high--velocity Fe~II
component, from 22,000 \kms\ to 29,000 \kms.  Note that this leaves a
gap from 17,000 to 22,000 where the adopted Fe~II optical depth is
zero.  Introducing the high--velocity component of Fe~II accounts
rather well for the 4300 and 4700~\AA\ absorptions (and it also
improves the 3900--4200~\AA\ region).

For later reference we mention here that the apparent blue edge of the
Ca~II H\&K absorption feature reaches to 40,000 \kms\ (or 32,000 \kms\
if formed by Si~II $\lambda$3858, or 27,000 \kms\ if formed by Si~III
$\lambda$3801); the blue edge of the red Si~II feature formed by
$\lambda$6355 reaches to 25,000 \kms.

\subsection{Eight and Two Days Before Maximum}

In the upper panel of Figure~2 an observed $-8$ day spectrum is
compared with a synthetic spectrum that has $v_{phot}$ = 12,000 \kms.
The other fitting parameters are listed in Table~2.  Here, we use only
a weak contribution from high--velocity Fe~II, from 20,000 to 25,000
\kms, and without it the fit would be only slightly worse.  But now,
in this synthetic spectrum, we use two components of Ca~II. A
high--velocity (25,000 to 40,000 \kms) component of Ca~II is the only
plausible way we have found to account for the observed absorption
from 7800 to 8000~\AA.  This feature is present in other spectra at
similar epochs and therefore is definitely a real feature; we
attribute it to the Ca~II infrared triplet, forming in the
high--velocity component.

In the lower panel of Figure~2 an observed $-2$ day spectrum is
compared with a synthetic spectrum that has $v_{phot} = 11,000$ \kms.
The other fitting parameters are listed in Table~3.  In the synthetic
spectrum we still are using two components of Ca~II, one extending
from the photosphere to 16,000 \kms, and the other from 20,000 to
23,000 \kms.  This two--component calcium leads to good agreement with
the observed ``split'' of the Ca~II H\&K feature.  Other ways to
account for the H\&K split might be to invoke Si~II $\lambda3858$ or
Si~III $\lambda3801$ (Kirshner et~al. 1993; H\"oflich 1995; Nugent
et~al. 1997; Lentz et~al. 1999) but at this epoch of SN~1994D we find
that in LTE, at least, other Si~II or Si~III lines would have to be
made too strong compared with the observations.  Our main reason for
preferring the two--component Ca~II at this epoch, however, is that it
can account for the observed absorption near 8000~\AA.  (As we will
discuss in a future paper, we find strong support for this
interpretation in Figure~4 of Meikle \& Hernandez (1999), which
compares spectra of SNe~1981B, 1994D, and 1998bu.  In SNe~1994D and
1998bu both the H\&K and the infrared--triplet absorption features are
split, while in SN~1981B neither is split.)

At present we have no explanation for the general difference between
the levels of the observed and synthetic spectra from 6500 to
7900~\AA, other than to note that we are inputting a simple blackbody
continuum from the photosphere.

The inset in the lower panel of Figure~2 compares the observed
spectrum near the 1.05 $\mu$m IR absorption to a synthetic spectrum
that is an extension of the optical synthetic spectrum.  We see that
at this epoch Mg~II, with the same reference--line optical depth we
use to fit the optical features, accounts nicely for the IR
absorption.  Thus we, like Wheeler et~al. (1998), favor the Mg~II
identification for the infrared absorption.  The inset also
illustrates the possibility that O~I may be affecting the spectrum
near 1.08 $\mu$m; here, in order not to mutilate the Mg~II feature, we
have had to reduce the O~I reference--line optical depth by a factor
of three compared to the value we used for the optical spectrum, which
we would have to attribute to NLTE effects.

\subsection{Evolution of the Ca~II H\&K Feature}

Figure~3 shows the evolution of the Ca~II H\&K feature.  If, as we
suspect, the whole profile at $-12$ and $-11$ days is dominated by
Ca~II, then {\sl some} Ca~II must be present throughout the interval
15,000 -- 40,000 \kms. At later times, from 21 to 74 days, the Ca~II
absorption forms only at velocities less than about 15,000 \kms.  (The
weak absorption near 3650 \AA\ is probably produced by an iron--peak
ion rather than by Ca~II.)  However, between $-9$ and $-3$ days the
profile undergoes a complex evolution.  While the highest--velocity
absorption ($>22,000$ \kms) fades away, and the low--velocity
absorption ($<15,000$ \kms) develops, a dip appears near 19,000 \kms\
and a peak develops near 16,000 \kms.  The 19,000 \kms\ dip is the
bluer component of the split discussed above for $-2$ days.  As
explained in the previous section we suspect that the 19,000 \kms\
minimum is caused by Ca~II (rather than by Si~II or Si~III) mainly
because of the Ca~II IR triplet.  If this is so, then at these phases
there must be a local minimum in the Ca~II radial optical depth
profile around 16,000 \kms.  This is not necessarily inconsistent with
the smooth H\&K profiles at $-12$ and $-11$ days if at those early
times the line optical depth was high throughout the 15,000 to 40,000
\kms\ interval.

\section{Discussion}

Assuming that our line identifications are correct, what is the cause
of this complex spectral behavior?  One part that seems clear is that
all of the matter detected at velocities lower than about 16,000 \kms\
represents freshly synthesized material.  And the highest--velocity
matter --- calcium up to 40,000 \kms, iron up to 30,000 \kms, and
silicon up to 25,000 \kms (from the blue edge of the red Si~II line),
is likely to be primordial.  This would be consistent with the
predictions of Hatano et~al. (1999), for a composition in which
hydrogen and helium have been burned to carbon and oxygen and the mass
fractions of heavier elements are just solar; in this case the ions
that are most likely to produce detectable spectral features from the
primordial abundances are just the three that do appear to be detected
at high velocity --- Ca~II, Fe~II, and Si~II.

However, what is the cause of the optical--depth minima of Ca~II and
Fe~II, somewhere around 16,000 \kms?  One possibility is that the
ionization and excitation structure is such that the Ca~II and Fe~II
lines, which decrease in strength with increasing temperature, have
optical--depth minima around 16,000 \kms.  Some weak support for this
comes from the fact that in the $-12$ day spectrum we found Fe~III
lines, forming just between 19,000 and 20,000 \kms, to be helpful in
fitting the spectrum.  In addition, in detailed atmosphere
calculations for model W7 with various primordial metallicities, Lentz
et~al. (1999) find a temperature maximum for low metallicity (see also
H\"oflich 1995).  SN~1994D had an unusually negative value of $U-B$
for a Type~Ia supernova, which could be an indication of low
metallicity.

Another possibility is that there is a local minimum in the radial
density profile around 16,000 \kms.  Delayed--detonation models have
smoothly decreasing densities in their outer layers, and deflagration
models tend to have only low--amplitude density peaks and
dips. Pulsating--detonation and tamped--detonation models have more
pronounced density minima, but at least in published models they occur
at velocities that are lower than 16,000 \kms\ (Khoklov, M\"uller, \&
H\"oflich 1993).

A third possibility is a nuclear explanation.  Detailed
nucleosynthesis calculations in delayed--detonation models recently
have been carried out by Iwamoto et~al. (1999).  As we will discuss
more thoroughly in another paper, the composition structures of some
of their models are generally consistent with most of the constraints
that we have inferred for SN~1994D.  In their model CS15DD1, for
example, the fractional abundances of freshly synthesized silicon,
sulfur, calcium, and iron begin to drop sharply above about 15,000
\kms, consistent with what we find for SN~1994D.  It is interesting
that in CS15DD1 the fractional abundances of sulfur and argon have
pronounced minima around 16,000 \kms.  Perhaps a delayed--detonation
model could be constructed to have a minimum in the fractional
abundance of freshly synthesized calcium around 16,000 \kms, although
it seems unlikely that such could be the case for iron.  It should be
noted that the synthesis of calcium and iron depends on the initial
metallicity (H\"oflich et~al. 1998).  Iwamoto et~al. did not include
primordial metals in their models.  Perhaps when primordial metals are
included, models will be found in which both calcium and iron have
fractional abundance minima near 16,000 \kms, if nuclear reactions can
reduce the levels of calcium and iron below the primordial levels at
this velocity.  These possibilities appear to be plausible and they
are being investigated (F.--K. Thielemann, personal communication).

\section{Conclusion}

Based on our interpretation of the pre--maximum--brightness spectra of
SN~1994D, it appears that given high--quality spectra obtained at
sufficiently early times, it should be possible to probe the
primordial composition of the SN~Ia progenitor (see also Lentz
et~al. 1999).  This will be important in connection with using
high--redshift SNe~Ia as distance indicators for cosmology (Branch
1998; Perlmutter et~al. 1998; Riess et~al. 1998), for testing the
predicted dependence of hydrodynamical models on primordial
composition (H\"oflich et~al 1998), and for testing
the prediction that low metallicity inhibits the ability of white
dwarfs to produce SNe~Ia (Kobayashi et~al. 1998).  It also appears
that early spectra of SNe~Ia can be used to place useful constraints
on the nature of the nuclear burning front in the outer layers of the
ejected matter.

Transforming the present somewhat qualitative indications into
reliable quantitative results will require (1) further parameterized
spectrum calculations, as part of a detailed comparative study of
early--time spectra of SNe~Ia (Hatano et~al., in preparation); (2)
detailed NLTE calculations for SN~Ia hydrodynamical models before the
time of maximum light (Nugent et~al. 1997; H\"oflich et~al. 1998;
Lentz et~al. 1999); (3) further detailed calculations of
nucleosynthesis in parameterized hydrodynamical explosion models
(Iwamoto et~al. 1999); and (4) many more high--quality observed
spectra of SNe~Ia before the time of maximum brightness.

\bigskip

We are grateful to Dean Richardson and Thomas Vaughan for assistance,
to Friedel Thielemann for discussions of nucleosynthesis in
hydrodynamical models, and to Ferdinando Patat and Peter Meikle for
making their observed spectra available in electronic form.  This work
was supported by NSF grants AST-9417102 to D.B., AST-9731450 to E.B.,
AST-9417213 to A.V.F., and NASA grant NAG5-3505 to E.B. and D.B.

\clearpage

\begin {references}

\reference{} Branch, D. 1998, ARAA, 36, 17

\reference{} Filippenko, A. V. 1997a, in Thermonuclear Supernovae,
ed. P. Ruiz--Lapuente, R. Canal, \& J. Isern, Kluwer, Dordrecht, p. 1

\reference{} Filippenko, A. V. 1997b, ARA\&A, 35, 309

\reference{} Fisher, A. 1999, Ph.D thesis, University of Oklahoma

\reference{} Fisher, A., Branch, D., Nugent, P., \& Baron, E. 1997,
ApJ, 481, L89

\reference{} Fisher, A., Branch, D., Hatano, K., \& Baron, E. 1999,
MNRAS, 304, 67

\reference{} Hatano, K., Branch, D., Fisher, A., Millard, J., \&
Baron, E. 1999, ApJS, in press (astro--ph/9809236)

\reference{} H\"oflich, P. 1995, ApJ, 443, 89

\reference{} H\"oflich, P., Wheeler, J. C., \& Thielemann,
F.--K. 1998, ApJ, 495, 617

\reference{} Iwamoto, K., Brachwitz, F., Nomoto, K., Kishimoto,~N.,
Hix,~W.~R., \& Thielemann,~F.--K. 1999, ApJ, submitted

\reference{} Khoklov, A., M\"uller, E., \& H\"oflich, P. 1993, A\&A,
270, 223

\reference{} Kirshner, R. P. et al. 1993, ApJ, 415, 589

\reference{} Kobayashi, C., Tsujimoto,~T., Nomoto,~K., Hachisu,~I., \&
Kato,~M. 1998, ApJ, 503, L155

\reference{} Lentz, E. J., Baron, E., Branch,~D., Hauschildt,~P.~H.,
\& Nugent,~P.~E. 1999, ApJ, submitted

\reference{} Mazzali, P. A., \& Lucy, L. B. 1998, MNRAS, 295, 428

\reference{} Meikle, W. P. S. et al. 1996, MNRAS, 281, 263

\reference{} Meikle, W. P. S. \& Hernandez, M. 1999,
J. Ital. Astr. Soc., in press

\reference{} Millard, J. et~al. 1999, ApJ, in press

\reference{} Nomoto, K., Thielemann, F.--K., \& Yokoi 1984, ApJ, 286, 644

\reference{} Nugent, P., Baron, E., Branch, D. Fisher, A., \&
Hauschildt,~P.~H. 1997, ApJ, 485, 812

\reference{} Patat, F., Benetti, S., Cappellaro, E., Danziger, I. J.,
Della Valle,~M., Mazzali,~P.~A., \& Turatto,~M. 1996, MNRAS, 278, 111

\reference{} Perlmutter, S. et~al. 1999, ApJ, in press

\reference{} Richmond, M. W. et al. 1995, AJ, 109, 2121

\reference{} Riess, A. G. et al. 1998, AJ, 116, 1009

\reference{} Treffers, R. R., Filippenko, A. V., Van Dyk, S. D., \&
Richmond, M. W. 1994, IAU Circ. No. 5946

\reference{} Vacca, W. D., \& Leibundgut, B. 1996, ApJ, 471, L37

\reference{} Wheeler, J. C., H\"oflich, P., Harkness, R. P., \&
Spyromilio,~J. 1998, ApJ, 496, 908

\end{references}

\clearpage

\begin{figure} 
\figcaption{A $-12$ day observed spectrum in $F_{\lambda}$ (Filippenko
1997a,b) is compared with synthetic spectra that have
$v_{phot}=15,000$ \kms.  In the upper panel, ions that contribute
features in the synthetic spectrum are labeled.  In the lower panel a
second, high--velocity component of Fe~II is added.  In this and
subsequent figures, the spectra are in the SN~1994D rest frame.}
\end{figure}

\begin{figure} 
\figcaption{In the upper panel, a $-8$ day observed spectrum in
$F_{\nu}$ (Patat et~al. 1996) is compared to a synthetic spectrum that
has $v_{phot}=12,000$ \kms.  In the lower panel, a $-2$ day observed
spectrum in $F_{\nu}$ (Patat et~al. 1996) is compared to a synthetic
spectrum that has $v_{phot}=11,000$ \kms.  In the inset of the lower
panel a $-2$ day near--infrared spectrum (Meikle et~al. 1996) is
compared with an extension of the optical synthetic spectrum, except
that the O~I reference--line optical depth has been reduced by a
factor of three.  (The narrow spike near 1.13~$\mu$m is an artifact.)}
\end{figure}

\begin{figure} 
\figcaption{Observed spectra in the region of the Ca~II H\&K feature
(Filippenko 1997a,b). The solid vertical line corresponds to 3945~\AA\
(the $gf$--weighted rest wavelength of Ca~II H\&K) and the dashed
vertical lines are blueshifted by 10,000, 20,000, and 30,000
km~s$^{-1}$.}
\end{figure}

\end{document}